\documentstyle[aps,prd,graphics]{revtex}

\def\sw {\sin \theta_W}

\def\cw {\cos \theta_W}
\def\zp {Z^{\prime}}

\begin{document}
\twocolumn[\hsize\textwidth\columnwidth\hsize\csname
@twocolumnfalse\endcsname 

\title{Implications of a new light gauge boson for neutrino physics}
\author{C\'eline B\oe hm$^1$}
\address{$^1$DWB, 1 Keble Road, OX1 3RH, Oxford, England, UK,
\\
boehm@astro.ox.ac.uk}
\date{\today} 
\maketitle

\begin{abstract}
We study the impact of light  gauge bosons 
on neutrino physics. We show that they can explain the NuTeV anomaly and 
also escape the constraints from neutrino experiments if 
they are very weakly coupled and have a mass of a few GeV.   
Lighter gauge bosons with stronger couplings could explain 
both the NuTeV anomaly and the positive anomalous magnetic moment of the muon.  
However, in the simple model we consider in this paper (say a purely vectorial extra $U(1)$ current), 
they appear to be in conflict with the precise measurements of $\nu_{\mu}-e$, $\nu_e-e$ 
elastic scattering cross sections.  The surprising agreement that we obtain  
between our naive model and the NuTeV anomaly for $m_{\zp} \sim$ GeV  may  be a coincidence. 
However, we think it is interesting enough to deserve attention and perhaps a more careful analysis, 
especially since a new light gauge boson is a very important ingredient for the Light Dark Matter scenario. 
\end{abstract}


]

\section{Introduction}

For decades, particle physics experimentalists have tested the Standard Model predictions in $e^+ e^-$ 
and $p \bar{p}$ accelerators. The impressive agreement between 
independent measurements of the $Z$ and $W$ masses now sets 
$\sw^2 = 1 - \frac{m_{W}^2}{m_Z^2} = 0.2227$ (on-shell)\cite{lep}, where $\theta_W$ is 
the mixing angle (also called the Weinberg angle) defined as: 
\begin{eqnarray}
Z^{\mu} &=& \cos \theta_W W_3^{\mu} - \sin \theta_W B^{\mu} \nonumber \\
A^{\mu} &=& \sin \theta_W W_3^{\mu} + \cos \theta_W B^{\mu}. \nonumber
\end{eqnarray} 
One would expect to find a similar result from low energy (notably neutrino) experiments. 
In fact, it is not so. 

For example, the NuTeV experiment reports a three standard deviations above the Standard Model value for $\sw^2$ 
\cite{nutev}. A large deviation of $\sw^2$ was also found by an experiment done in 1976 which measured the 
$\bar{\nu}_e e$ elastic scattering cross section \cite{reines}. And even with the precise measurement of the 
$\nu_e e$ elastic scattering cross section by the LSND experiment (at the Los Alamos Neutron Science Center) \cite{lsnd} 
and the $\nu_{\mu} e$, $\bar{\nu}_{\mu} e$ elastic scattering processes 
by CHARM II at CERN \cite{charmii}, there is still room for non-standard physics. 

Other tests of the Standard Model seem to go in the same direction. Atomic parity violation experiments 
have obtained a better fit of their data by taking a value of $\sw^2$ that is slightly lower than the 
theoretical prediction at low energy \cite{apv}. The $Z$ decay width seems also lower than expected and, finally, 
a positive anomalous magnetic moment for the muon has been reported recently by the E821 experiment \cite{E821}.
Their measurement is above the Standard Model prediction by 2.7 standard deviations. 
Although Supersymmetry could explain this deviation, we shall see that a light gauge boson could do the same job. 
(Reciprocally, this measurement sets the maximal couplings a new 
gauge boson can have.) It is then tempting to see whether a gauge boson with these couplings 
could explain the NuTeV anomaly and be compatible with precision measurements. 

Extra gauge bosons were introduced a long time ago, notably in the context of $SO(10)$ 
or $E(6)$ grand unified theories. The implications of heavy a $\zp$ on \textit{eg}  
neutrino physics, the NuTeV anomaly and atomic parity violation  have been studied at length   
\cite{langacker}. Light/very light gauge bosons  were also proposed  in the past,  
in the context of supersymmetric theories \cite{pierreu1} and later to explain the NuTeV anomaly \cite{sacha}.  
Here, we shall just focus on masses of about a few hundred MeV to a few GeV. As we shall see, they appear to be the most 
interesting for our purpose. We note that our results for a gauge boson of a few GeV seem to confirm previous findings 
\cite{sacha}, although we do not consider the same couplings.

A light gauge boson turns out to be a crucial element for the Light Dark Matter scenario as it allows one to obtain 
an annihilation cross section proportional to the square of the 
Dark Matter velocity (which is necessary to not overproduce gamma rays in the galactic centre   
and simultaneously achieve the correct relic density) \cite{bens,bf}. In fact, there may be other ways to obtain 
an annihilation cross section with the desired property. However, if a light gauge boson 
was discovered, this could be quite in favour of the Light Dark Matter scenario (depending on the mass and the 
couplings of this gauge boson). If no new spin$-1$ particles are discovered, one could instead constrain the 
Light Dark Matter (LDM) parameter space. Alternatively, one may have to consider a light gauge boson with 
no or extremely small couplings to neutrinos, which in fact may help nucleosynthesis, as 
was found in \cite{raffelt}.

 Here, we do not propose a theory or mechanism that could be responsible for a light gauge particle. 
 Our aim is rather to exhibit unexpected properties of light gauge bosons that may perhaps 
 deserve more attention. More theoretical/particle physics arguments can however be found in \cite{pierreu1,bf}. 
 In fact, if the features mentioned in this paper are correct, then there is still 
 room for extra physics at low energy. Alternatively, this could mean that we have already detected it. Whether 
 this is the case or not, the properties exhibited in this paper could maybe give additional motivations to 
 neutrino experiments to measure even more precisely the $\nu_{e, \mu} e$ and $\bar{\nu}_{e,\mu} e$ elastic scattering cross sections 
 at low energy since this is crucial to conclude about the existence of light gauge bosons.

\section{Fitting Standard Model deviations with a light gauge boson}

\subsection{Muon anomalous magnetic moment}
Although most of the Standard Model features have been confirmed experimentally with extremely 
impressive accuracy, a few precision measurements seem to disagree with the 
Standard Model predictions at two or three standard deviations. For example,  
the E821 collaboration at Brookhaven Alternative Gradient Synchroton found, in a recent analysis, 
that the experimental value of the anomalous magnetic moment ($g-2$) for the negative muon 
differs from the latest theoretical value (where the hadronic contribution is determined directly 
by using the data from electron-positron collisions) by 2.8 standard deviations 
\cite{E821}. 

\vspace{0.3cm}
Combined with their previous measurement of the anomalous magnetic moment of the positive 
muon, E821 found that the experimental average for the muon $g-2$ differs from the theoretical value by 
2.7 standard deviations. (This deviation is only 1.4 standard deviations by using the  
indirect determination, say with the decay of taus into hadrons). 
The discrepancy between the Standard Model \cite{gm2th} and the average experimental value is about: 
$\delta a_{\mu} = a_{exp} - a_{th} \sim  (2.7 \pm 1.04) \ 10^{-9} $ (using $e^+ e^-$ data) and 
$\delta a_{\mu} = a_{exp} - a_{th} \sim  (1.2 \pm 0.92) \ 10^{-9} $ (using tau data). 

\vspace{0.3cm}
Such a deviation could be the signature of new physics. For example, this could point 
towards Supersymmetry \cite{g2susy}. However, as long as unequivocal signatures are not found, 
it is premature to conclude that supersymmetry is responsible for 
the positive $g-2$ found by E821, especially if other particles could provide the same signature.

\vspace{0.3cm}
We shall see that a new (neutral) light gauge boson could give a positive 
contribution to the muon $g-2$ (as already found in \cite{bf}) and, simultaneously, explain the NuTeV 
anomaly but this also implies a too large (and unseen) contribution to the 
$\nu_{l} e$ and $\bar{\nu}_{l} e$ elastic scattering cross sections  
(at least if the associated extra $U(1)$ current is purely vectorial).

\vspace{0.3cm}
We shall write the $\zp$ coupling to ordinary fermions ($\zp-\bar{f}-f$) 
as $[i \gamma^{\mu} \, (u_l P_l + u_r P_r)]$ with $u_l = u_r$ to avoid axial contribution and  
$P_r$, $P_l$ the chiral projectors. The factor $i$, in the above expression, is set 
to match the conventions that we think have been taken in the literature.  
According to the formula in \cite{leveille} (where we take the relevant limits): 
\begin{itemize}
\item 
The extra contribution of a new gauge boson, if it is heavier than the muon, is given by:
$$\delta a_{\mu} \sim \frac{u_{l}^2}{12 \pi^2} \frac{m_{\mu}^2}{m_{\zp}^2}.$$
We note that our formula differs from that displayed in \cite{sacha}. This is because we wrote  
our couplings in a different way. Using the same conventions as they used, say $u_l = 9 g_{\zp}$, 
we obtain the same formula. 

This actually sets a constraint on the maximal coupling that an extra gauge boson can have: 
$$ u_l^{\mu} \sim 3.28 \ 10^{-6} \left(\frac{m_{\zp}}{\rm{MeV}}\right) 
\ \left(\frac{\delta a_{\mu}}{10^{-9}}\right)^{1/2}.$$ 

\item If the extra gauge boson is lighter than the muon, then the extra contribution is  given by 
$$\delta a_{\mu} \sim \frac{u_{l}^2}{8 \pi^2}.$$ A very light gauge boson with couplings of about  
$$ u_l^{\mu} \sim 2.81 \ 10^{-4}  \ \left(\frac{\delta a_{\mu}}{10^{-9}}\right)^{1/2}$$
could then possibly fit the experimental value of the muon $g-2$. 
\end{itemize}

\vspace{0.3cm}
This extra gauge boson would also contribute to the electron anomalous magnetic moment. 
The latter is measured at a better level of accuracy so one has to check that the $\zp$ 
contribution to the electron $g-2$ is not experimentally forbidden. 
For a $\zp$ heavier than the electron, one finds 
$$\delta a_{e} \sim \frac{u_{l}^2}{12 \pi^2} \frac{m_{e}^2}{m_{\zp}^2}.$$ 
This sets the $\zp$ coupling to 
electrons to be at most:
$$ u_l^e = 2.17 \ 10^{-5} \ \left(\frac{m_{\zp}}{\rm{MeV}}\right) 
\left(\frac{\delta a_{e}}{10^{-12}}\right)^{1/2}.$$ 

\vspace{0.3cm}
Let us make a ``universality'' assumption, say $|u_l^e|=|u_l^{\mu}|$. 
Since these two couplings strongly depend on the measured 
electron and muon $g-2$ respectively, it is not completely trivial to know whether $|u_l^e|$ 
must take the value of $|u_l^{\mu}|$ or vice versa. In fact, 
the answer depends on the mass of the new gauge boson.  

\vspace{0.4cm}
$\star$
For $m_{\zp} > m_{\mu}$: One cannot impose $|u_l^{\mu}|$ to be equal to the maximal value of $|u_l^e|$ as the $\zp$ would 
yield a too large contribution to the muon $g-2$. On the other hand, one can safely impose 
$|u_l^e|$ to be equal to 
$$u_l^e \sim 3.28 \ 10^{-6} \ \left(\frac{m_{\zp}}{\rm{MeV}} \right)
\left(\frac{\delta a_{\mu}}{10^{-9}}\right)^{1/2}.$$
This yields the following prediction for the electron $g-2$: 
$$\left(\frac{\delta a_{e}}{10^{-12}}\right) \sim 2.285 \, 10^{-2} \, \left(\frac{\delta a_{\mu}}{10^{-9}}\right). $$

\vspace{0.3cm}
$\star$
For $m_{\zp} < m_{\mu}$: both values for $|u_l^{\mu}|$ and $|u_l^e|$ 
can be accommodated if $m_{\zp} \sim 12.95 \, (\frac{\delta a_{\mu}}{10^{-9}})^{1/2} \, (\frac{\delta a_{e}}{10^{-12}})^{-1/2} $ MeV. 
In fact, equating the two couplings provides the following relationship:
$$\left(\frac{\delta a_{\mu}}{10^{-9}}\right) \sim 5.963 \ 10^{-3} \ \left(\frac{m_{\zp}}{\rm{MeV}}\right)^2  
\ \left(\frac{\delta a_{e}}{10^{-12}}\right).$$

\vspace{0.4cm}
If a light gauge boson with a mass $m_{\zp} < m_{\mu}$ turns out to be responsible for the muon $g-2$, 
then there should be a contribution to the electron $g-2$ that should be testable in the near future. 

\vspace{0.3cm} 
Let us now investigate the effect of a light neutral (and weakly coupled) gauge boson on neutrino physics.

\subsection{NuTeV anomaly}
As mentioned in the introduction,  a ``famous'' deviation,  
that led to many theoretical investigations,  
comes from NuTeV experiment which measures three standard deviations above the Standard Model 
predictions for the value of $\sw^2$. Standard physics explanations, such as 
electroweak radiative corrections, QCD corrections (perturbative QCD, charged current charm production, 
parton distribution functions, isospin breaking, nuclear effects), as well as experimental effects 
have been addressed \cite{mcfarland,kretzer}. An asymmetric strange 
sea and/or an isospin violation in parton distributions functions could be two possible 
explanations \cite{sacha,kretzer} but the uncertainties are still too large to draw definite conclusions. 
In fact, in some cases, they could even enhance the discrepancy with the Standard Model. 
 
\vspace{0.3cm}  
Yet, it may still be legitimate to look for non-standard explanations. The existence of a new gauge boson 
has been already proposed in the context of this anomaly (see notably \cite{sacha}). It was seen that either a 
light (with $ 1 < m_{\zp}< 10 $ GeV) or heavy gauge boson could be an explanation. This analysis was specifically 
addressed for solving the NuTeV anomaly and the limits from neutrino scattering experiments were not discussed 
in the case of a light gauge boson.  In contrast, the new gauge boson we consider is related to the Light Dark Matter scenario 
and we pay very much attention to neutrino experiments as they directly and indirectly constrain the couplings of this 
new $\zp$ to ordinary matter and Dark Matter. Despite the difference in our motivations, we confirm that a $\sim$ GeV gauge boson 
could explain the NuTeV anomaly. However, we note that our couplings seem smaller than those mentioned in \cite{sacha}.

\vspace{0.3cm} 
NuTeV is an experiment measuring the ratio $$R= \frac{\mbox{neutral currents}}{\mbox{charged currents}} 
= (g_l^2 - g_r^2),$$ with $g_{l,r}^2= [(g_{l,r}^u)^2+ (g_{l,r}^d)^2]/4$ and where $g_{l,r}^f = 2(T_3(f_{l,r}) - Q(f) \sw^2)$  
are the left and right couplings of the $Z$ boson to fermions that we shall write, in a more conventional way, as:
$$ Z_l P_l + Z_r P_r = \frac{-i g}{2 \cw} (g_l^f P_l + g_r^f P_r) = \frac{-i g}{2 \cw} (c_v - c_a \gamma_5).$$  
More precisely, NuTeV measured (according to the Paschos-Wolfenstein suggestion \cite{paschos}) the observable:
$$R = \frac{\sigma_{(\nu_{\mu} N\rightarrow \nu_{\mu} X)} - \sigma_{(\bar{\nu}_{\mu} N \rightarrow \bar{\nu}_{\mu} X)}}{
\sigma_{(\nu_{\mu} N\rightarrow \mu X)} - \sigma_{(\bar{\nu}_{\mu} N \rightarrow \mu^+ X)}}.$$

\vspace{0.3cm} 
A new neutral gauge boson will contribute both to $\sigma_{(\nu_{\mu} N\rightarrow \nu_{\mu} X)}$ and 
$\sigma_{(\bar{\nu}_{\mu} N \rightarrow \bar{\nu}_{\mu} X)}$. 
The square amplitude for the $\nu_{\mu}-q$ elastic process is given by: 
\begin{eqnarray}
|M|^2 &=& |M_Z|^2 + |M_{\zp}|^2  + 2 |M_{\zp} M_{Z}^{\star}|.\nonumber
\end{eqnarray}
with  $|M_{\zp}|^2$ and $2 |M_{\zp} M_{Z}^{\star}|$ the 
additional contributions due to the presence of a $\zp$.  
The denominator of the NuTeV observable 
($R$) is related to charged currents. It is not affected by the presence 
of a new neutral gauge boson. 

\vspace{0.3cm}
In the following, we shall assume that this $\zp$ has no axial coupling (say $u_l = u_r$ or $c_a^{\zp} =0$), as 
initially considered in the Light Dark Matter scenario \cite{bf}. 
The assumption of a purely vectorial extra $U(1)$ current could be alleviated but not 
without generating problems that would spoil the simplicity of our model.

\vspace{0.3cm}
We shall consider a gauge boson light enough to satisfy the relationship 
$m_{\zp}^2 < Q^2_{NuTeV}$ (with $Q^2_{NuTeV} \in [16,25]$ GeV$^2$,       
the transfer momentum usually defined as $t=q^2=-Q^2$).  
Summing all the contributions (and with $c_v^{\nu} = 1/2$), we obtain that 
the square amplitude associated with the elastic scattering process 
$\nu_1 q_1 \rightarrow \nu_2 q_2$ (in presence of a new $\zp$) is given by: 
\begin{eqnarray}
|M|^2&=&  8 \ G_F^2 \ \ \left\{ (c_v + c_a)^2 p_{\nu 1}.p_{q 1} \ 
p_{\nu 2}.p_{q 2} \  \right. \nonumber \\
&& \left.  \hspace{1.2cm} + (c_v - c_a)^2 \ p_{\nu 1}.p_{q 2} \  p_{\nu 2}.p_{q 1} \right\} \nonumber\\
&+&  \frac{8 \, u_l^{\nu \, 2} \,  u_l^{q \, 2}}{Q^4}  
\left\{p_{\nu 1}.p_{q 1} \ p_{\nu 2}.p_{q 2} + 
\, p_{\nu 1}.p_{q 2} \ p_{\nu 2}.p_{q 1} \right\} \nonumber \\
&+&  \left(\frac{16 G_F }{\sqrt{2} \ Q^2}\right) \ u_l^{\nu} \  u_l^{q } \nonumber \\
&&\left\{ (c_v + c_a) \ p_{\nu 1}.p_{q 1} \ p_{\nu 2}.p_{q 2} \  \right. \nonumber \\
&& \left. \hspace{2.5cm} + \, (c_v - c_a)  \ p_{\nu 1}.p_{q 2} \ p_{\nu 2}.p_{q 1}\right\}, \nonumber
 \end{eqnarray}
where $m_{\zp}^2 \ll Q^2 \ll m_Z^2$.
 
 \vspace{0.3cm}
Let us now compute the square amplitude associated with the $\bar{\nu}_{\mu}-q$ elastic scattering cross section. 
For the Standard Model, it has been shown that the only difference is the permutation of 
$p_{\nu 1}.p_{e 1} \ p_{\nu 2}.p_{e 2}$ and  $p_{\nu 1}.p_{e 2} \ p_{\nu 2}.p_{e 1}$. Thus, we obtain:
\begin{eqnarray}
|M|^2_{\bar{\nu}_{\mu} N \rightarrow \bar{\nu}_{\mu} X} &=&  8 \ G_F^2 \ \left\{ (c_v + c_a)^2 p_{\nu 1}.p_{q 2} 
\ p_{\nu 2}.p_{q 1} \ \right. \nonumber \\  
&& \left. \hspace{1.2cm} + \ (c_v - c_a)^2 \ p_{\nu 1}.p_{q 1} \  p_{\nu 2}.p_{q 2} \right\}   \nonumber \\ 
&+& 8 \ \frac{u_l^{\nu \ 2} \  u_l^{q \ 2}}{Q^4}  \ 
\left\{p_{\nu 1}.p_{q 2} \ p_{\nu 2}.p_{q 1} \ \right. \nonumber \\  
&& \left. \hspace{2.3cm} + \ p_{\nu 1}.p_{q 1} \ p_{\nu 2}.p_{q 2} \right\}  \nonumber \\
&+&  \left(\frac{16 G_F }{\sqrt{2} \, Q^2}\right) u_l^{\nu} \,  u_l^{q}  \nonumber \\
&&\left\{ (c_v + c_a)  p_{\nu 1}.p_{q 2} \, p_{\nu 2}.p_{q 1} \right. \nonumber \\
&& \left. \hspace{1.5cm} + (c_v - c_a)  \ p_{\nu 1}.p_{q 1} \ p_{\nu 2}.p_{q 2}\right\} \nonumber
 \end{eqnarray}

\vspace{0.3cm}
With this new gauge boson, the numerator of the observable $R$ is then proportional to: 
\begin{eqnarray}
R_{num}&=&   \sum_{u,d} \left[ G_F  c_v^q c_a^q +  \frac{1}{\sqrt{2}}  
\left(\frac{u_l^{\nu} \, (c_a^q u_l^{q})}{\ Q^2}\right)  \right]\nonumber
\end{eqnarray}
where we summed on the quarks u and d, as required for isoscalar targets (as considered by NuTeV) and 
defined $R_{num}$ as $R_{num} = R/A$ with $A$ a constant that is not important for our purpose. 
Since $c_a^u = - c_a^d$, the contribution of the extra gauge boson vanishes, unless $u_l^u = - u_l^d$. 
Yet,  the $\zp$ contribution to $R_{num}$ is negative 
when $u_l^u < 0$ and $u_l^d >0$. It is positive when $u_l^u > 0$ and $u_l^d <0$. 

\vspace{0.3cm}
Strictly speaking, this condition indicates that one should relax the quark universality. 
However, to restrict our parameter space, we shall maintain  
$|u_l^u| = |u_l^d|$.

\vspace{0.3cm}
The NuTeV collaboration obtains a good fit of their data by taking 
\begin{eqnarray}
\sw^{2 \ on-shell} &=& 0.2277 \pm 0.0013 (stat) \pm 0.0009 (syst) \nonumber \\
&& - 0.00022 \left( \frac{M_{top}^2 - (175 \mbox{GeV})^2}{(50 \mbox{GeV})^2}\right)\nonumber \\
&& + 0.00032 \ln \, \left( \frac{M_{Higgs}}{150 \mbox{GeV}}\right). \nonumber
\end{eqnarray}
which corresponds (disregarding the theoretical uncertainties) to 
$G_F \sum_{u,d} c_v^q c_a^q = (3.1859 \pm 0.0257) \, 10^{-6}$, without radiative corrections 
(while one would expect 
$G_F \sum_{u,d} c_v^q c_a^q = 3.2444 \ 10^{-6}$ with $\sw^2 = 0.2227$).
If this anomaly is due to the presence of new physics, then the non standard contribution 
ro $R_{num}$ should be negative.  

\vspace{0.3cm}
To do a more accurate analysis, we shall take into account the radiative corrections. 
The latter affects the couplings $g_l$ and $g_r$ as follow:
\begin{eqnarray}
g_l &=& \, \sqrt{\rho} \ (T_3^f - Q \, (\sin^2 \theta)_{eff})\nonumber \\
g_r &=& - \, \sqrt{\rho} \  Q \, (\sin^2 \theta)_{eff}\nonumber 
\end{eqnarray}
where $\rho \equiv \rho_0=1$ without radiative corrections. Note that we used NuTeV conventions, 
in the above definition of the couplings. (With our conventions, $g_l^f = 2 g_l$ and $g_r^f = 2 g_r$.)
With their values of the couplings: $(g_l^2)^{NuTeV}=0.30005 \pm 0.00137$ and 
$(g_r^2)^{NuTeV}=0.03076 \pm 0.00110$ \cite{mcfarland}), 
we find 
$$(\sin^2 \theta)_{eff}^{NuTeV} = 0.2339^{+ 0.0036}_{-0.0037}$$ 
and 
$$\rho^{NuTeV} = 1.0120^{+ 0.0045}_{-0.0047}.$$ 
Those values can now be plugged into the expressions of $c_a$ and $c_v$ (still using NuTeV conventions):
\begin{eqnarray}
c_v &=& \sqrt{\rho} \ [(T_3^f - 2 \ Q \, (\sin^2 \theta)_{eff})]/2\nonumber \\
c_a &=& \sqrt{\rho} \ T_3^f/2, \nonumber 
\end{eqnarray}
to compute more accurately $G_F \sum_{u,d} c_v^q c_a^q$. 
We repeat the same procedure for the best fit of the Standard Model that NuTeV used for making its analysis  
(say $(g_l^{SM})_{eff}^2 =0.3042$ and $(g_r^{SM})_{eff}^2= 0.0301$). 
We find $(\sin^2 \theta)_{eff}^{SMBF} = 0.2307$ and $\rho=1.0179$. 

\vspace{0.3cm}
With the NuTeV values, we obtain: 
$$G_F \sum_{u,d} c_v^q c_a^q = (3.1507 \, \pm \, 0.0288 ) \ 10^{-6}$$ (while the Standard Model expectation is 
$G_F \sum_{u,d} c_v^q c_a^q =  3.2072 \ 10^{-6}$).

\vspace{0.3cm}
Assuming ``quasi'' universality, say  $u_l^u = - u_l^d$ 
and respecting the maximal value allowed by the muon $g-2$ (for the numerical 
example shown below: $\delta a_{\mu} = 1.5 \, 10^{-9}$), we find that 
a gauge boson of $\sim 314$ MeV can impressively explain the NuTeV anomaly (imposing 
$u_l^{d}=u_l^{\nu} = - u_l^u$). More precisely, using our previous expression, we obtain 
(for $Q^2 = 20$ GeV$^2$) that a gauge boson with a mass of $m_{\zp} = 314$ MeV, gives:  
$$R_{num.rel} = 3.1507 \, 10^{-6}$$ 
while NuTeV best fits gives 
$$R_{num.rel}^{NuTeV} = (3.1507 \pm 0.0288) \ 10^{-6}.$$ 
We note that our result is 
very sensitive to the $\rho$ parameter.

\vspace{0.3cm}
Other ranges of masses can also explain the NuTeV anomaly if the couplings are smaller. For example, 
if $u_l^{\nu} = 1.9 \ \sqrt{1.5} \ 10^{-6} \ (\frac{m_{\zp}}{\mbox{MeV}})$ (where $\delta a_{\mu} = 1.5 \ 10^{-9}$), 
then the best fit is obtained for $m_{\zp} = 544.5$ MeV (with $R_{num.rel} = 3.1507 \ 10^{-6}$).
If $u_l^{\nu} = 1.1 \ \sqrt{1.5} \ 10^{-6} \ (\frac{m_{\zp}}{\mbox{MeV}})$, 
then the best fit is obtained for $m_{\zp} = 955$ MeV (with $R_{num.rel} = 3.1508 \ 10^{-6}$).
This new gauge boson would then contribute to the positive muon $g-2$ without explaining it completely.

\vspace{0.3cm}
Another test to make sure that a light gauge boson could indeed solve the NuTeV anomaly is the 
precise value of the couplings. NuTeV finds (for $Q^2=20$ GeV$^2$):
$$(g_l^{eff})^2 =0.3001 \pm 0.0014$$ and 
$$(g_r^{eff})^2= 0.0308 \pm 0.0011$$ 
while the Standard Model expectations are 
$(g_l^{eff})_{SM}^2 =0.3042$ and $(g_r^{eff})_{SM}^2= 0.0301.$  
The discrepancy between NuTeV and the Standard Model best fit is therefore:
$$\Delta (g_l^{eff})^2 = - 0.0041 \pm 0.0014$$ and 
$$\Delta (g_r^{eff})^2= 0.0007 \pm 0.0011.$$ 

\vspace{0.3cm}
The left coupling would therefore 
decrease while the right one would increase. $g_l$ is associated to the scalar product 
$p_{\nu 1}.p_{q 1} \ p_{\nu 2}.p_{q 2}$ and $g_r$ to $ p_{\nu 1}.p_{q 2} \, p_{\nu 2}.p_{q 1}$. 
Without new particles, they are given by $\sum_{u,d}  (T_3^q -  q \sw^2)^2 = 1/2 - \sw^2 + 5/9 \sw^4$ 
and $\sum_{u,d} q^2 \sw^4 = 5/9 \sw^4$ respectively. A new gauge boson will change these couplings by the amount 
$\Delta (g_{l,r}^{\zp})^2$.

\vspace{0.3cm}
For $u_l^{\nu} = 1.1 \, \sqrt{1.5} \ 10^{-6} \, (\frac{m_{\zp}}{\mbox{MeV}})$, 
$u_l^{\nu} = u_l^{d} = - u_l^{u}$ and $m_{\zp} = 955$ MeV, we find: 
$$\Delta (g_l^{\zp})^2 = -0.0037$$ and $$\Delta (g_r^{\zp})^2 =0.0011.$$  
For $u_l^{\nu} = 3.28 \, \sqrt{1.5} \ 10^{-6} \, (\frac{m_{\zp}}{\mbox{MeV}})$, 
$u_l^{\nu} = u_l^{d} = - u_l^{u}$, $m_{\zp} = 314$ MeV, we obtain: 
$$\Delta (g_l^{\zp})^2 = -0.0037$$ and $$\Delta (g_r^{\zp})^2 = 0.0011.$$ 
Both exemples are in good agreement with NuTeV.

\vspace{0.3cm}
So far, we studied the case of "universal" couplings. 
However, it would make sense to relax this assumption (at least in the quark sector), 
as one would obtain a situation which would appear closer to the Standard Model. We choose not to do it because this 
would add more freedom to our study and somehow lessen the case for a light gauge boson. Indeed, even by maintaining "universality", 
we shall see that one can fit other constraints and this anomaly simultaneously. 

\vspace{0.3cm}
Note that it is possible to obtain a destructive interference term that could reduce the left-handed 
coupling enough to explain the NuTeV anomaly with very heavy gauge boson. 
This works for $m_{\zp} \sim $1-1.5 TeV. However, colliders cannot probe 
such a scenario yet. In contrast, we consider a mass range that was extensively 
investigated in accelerators but the $\zp$ that we introduce 
is likely to escape accelerator searches because of the smallness of its couplings. 
Also, it should not lead to large radiative corrections so this should not affect the precise 
measurement of $M_W$ in colliders. A more precise study is needed, but in absence 
of a given model, it seems difficult to get accurate estimates.

\section{Neutrino experiments} 
In the two previous sections, we saw that a light gauge boson which would be equally coupled  
to quarks and leptons could explain both the experimental value for the muon $g-2$ and the NuTeV anomaly. 
We shall now investigate the implications for the $\nu_{e,\mu} e$ elastic scattering cross sections at low energy.

\subsection{CHARM II experiment}

$\nu_{\mu} e \rightarrow \nu_{\mu} e$ and $\bar{\nu}_{\mu} e \rightarrow \bar{\nu}_{\mu} e$ 
scattering processes were observed by CHARM II experiment (at CERN) from 1987 till 1991. They found a good agreement with LEP 
results although $\sin^2 \theta_{eff}$ was determined with an uncertainty of 3.57 $\%$ \cite{charmii}: 
$$\sin^2 \theta_{eff}^{CHARM II} = 0.2324 \pm 0.0083.$$ 
They also determine $c_a$ and $c_v$ from the absolute 
$\nu_e$ scattering event rate. They found $c_v^{\nu e} = -0.035 \pm 0.017$ and 
$c_a^{\nu e} = -0.503 \pm 0.017$. More precisely, they quoted: $c_v^{\nu e} = -0.035 \pm 0.012 (stat) \pm 0.012 (syst)$ and 
$c_a^{\nu e} = -0.503 \pm 0.006 (stat) \pm 0.016 (syst)$ (with $c_v^{\nu e} = c_v^{\nu} c_v^e$ and 
$c_a^{\nu e} = c_a^{\nu} c_a^e$), which 
allows us to determine $\rho$ and $\sin^2 \theta_{eff}$.
This  reduces the uncertainties on the two cross sections significantly enough to restrict our parameter 
space. However, as we shall see, the CHARM II  mean value slightly differs from the 
Standard Model expectation.

\vspace{0.3cm}
$\sigma_{\nu_{\mu} e \rightarrow \nu_{\mu} e}$ 
can be obtained from the $\nu_{\mu}-q$ elastic scattering cross section by 
replacing the quark by an electron and changing $Q^2$ into $M_{\zp}^2$ (since this experiment 
takes place at low $Q^2$). 

\vspace{0.3cm}
The mean cross section measured by CHARM II is slightly smaller than 
the Standard Model prediction at low energy (a few percent of difference, although the Standard Model 
expectations is within the error bars). The introduction of a new light gauge boson can then 
help to reach a better agreement, provided the latter has very weak couplings (much smaller than those considered 
for explaining the $g-2$). We can fit the mean value of CHARM II findings by imposing 
$u_l^{\nu} \sim u_l^e \sim [0.3,0.6] \ \sqrt{1.5} \ 10^{-6} (m_{\zp}/\mbox{MeV})$  
(no significant deviation from the Standard Model prediction should be detected if 
$u_l^{\nu} = u_l^e < 0.1 \ \sqrt{1.5} \ 10^{-6} (m_{\zp}/\mbox{MeV})$). 

\vspace{0.3cm}
Our results depend on the sign of $u_l^e$. However for a coupling as small as $|u_l|^{\nu} \sim 0.1 \ \sqrt{1.5} \ 10^{-6} (m_{\zp}/\mbox{MeV})$,
having $u_l^e > 0$ or $u_l^e < 0$ does not make a big difference. On the other hand, 
it does when $|u_l|^{\nu} \sim [0.1,0.6] \ 10^{-6} (m_{\zp}/\mbox{MeV})$: 
the $\nu_{\mu} e$ cross section is lower when $u_l^e$ and $u_l^{\nu}$ have the same sign; 
it is larger otherwise.

\vspace{0.3cm}
The contribution of such a gauge boson 
to the muon anomalous magnetic moment and scattering processes would be negligible.  
Nevertheless, it would still explain the NuTeV anomaly if $m_{\zp} \gtrsim $ 1 GeV. 
One could relax universality or introduce axial couplings, say $c_a^{\zp} \neq 0$. 
The latter solution would presumably generate problems. However, the couplings mentioned above would be 
closer to those needed to explain the positive muon $g-2$.  
We will study this possibility in a forthcoming paper.

\vspace{0.3cm}
It is worth noting that the presence of other particles could nevertheless explain the muon $g-2$ measurement. 
One may find the scenario of a gauge boson supplemented by new fermions or new scalars, artificial. 
But these features are somehow already implemented in the Light Dark Matter framework \cite{bf}. 

\vspace{0.3cm}
If the Light Dark Matter candidate is a scalar that exchanges  
heavy fermions $F$, then the coupling of the LDM particles to muons and $F$ particles 
is expected to be about $c_s = 3.87 \, 10^{-5} \, (m_F/\mbox{MeV})^{1/2} (\delta a_{\mu}^F/10^{-9})$, in order 
to explain the muon $g-2$ discrepancy. Surprisingly enough, this value seems also to 
satisfy the requirement that the Dark Matter annihilation cross section does not exceed 
$10^{-31}$  cm$^3$ s$^{-1}$ \cite{bens}. In fact, with such couplings, we obtain 
$\sigma_{ann} v \sim 10^{-31}$ cm$^3$ s$^{-1}$. The $\zp$ exchange could then allow one to obtain 
the correct relic density but the positron line would be the signature of the $F$ exchange. 
Such a scheme would in addition explain the positive (experimental) value of the muon $g-2$. 
The $F$ particles could have escaped past neutrino 
or accelerator experiments if they are heavier than $\sim$100 GeV. But this point deserves a more careful study. 
However, if it is correct, then these extra particles are likely to be found in LHC since their couplings are 
not very suppressed.  Here, we suppose no pseudoscalar contribution but this could be relaxed.


\subsection{LSND experiment}

We saw that CHARM II experiment was setting stringent constraints on the couplings of a light $\zp$.
We can now determine whether the $\nu_{e} e$ elastic scattering cross section in presence of this new gauge boson  
is  compatible with the LSND experiment.
 
\vspace{0.3cm}
The $\nu_e e$ elastic scattering cross section can be almost inferred from the $\nu_{\mu}-e$ cross section, 
although one has to add the $W$ exchange. The LSND experiment at the Los Alamos Neutron Science Center has 
measured this elastic scattering cross section and found a good agreement with the  
Standard Model value. Strictly speaking, however, there is still room for non standard physics. 
The Standard Model value for the $\nu_e e$ elastic scattering cross section is (retaining the electron mass) 
$$ \sigma_{\nu-e}^{sm} = 9.3 \, 10^{-45} \left(\frac{E}{\rm{MeV}}\right) \rm{cm}^2.$$
LSND found: 
$$ \sigma_{\nu-e}^{exp} = 10.1 \pm 1.1 (stat) \pm 1.0 (syst) \, \left(\frac{E_{\nu e}}{\mbox{MeV}}\right) \, 10^{-45}  \rm{cm}^2.$$

\vspace{0.3cm}
For LSND, the transfer momentum $Q$ is much smaller than $m_{\zp}$. Therefore the cross section associated with 
a light gauge boson of a few hundred MeV to a few GeV could be considered in the limit $Q<<M_{Z^{\prime}}$.   
The extended square amplitude of the $\nu_e e$ elastic scattering process is now given by
\begin{eqnarray}
|M|^2 &=&  8 \ G_F^2 \ \left\{ (c_v' + c_a')^2 p_{\nu 1}.p_{e 1} \ p_{\nu 2}.p_{e 2} \ \right. \nonumber \\ 
&& \left. \hspace{1cm} +  \ (c_v' - c_a')^2 \ p_{\nu 1}.p_{e 2} \  p_{\nu 2}.p_{e 1} \right\}   \nonumber \\ 
&+& 8 \ \frac{u_l^{\nu \ 2} \  u_l^{e \ 2}}{M_{\zp}^4}  \ 
\left\{p_{\nu 1}.p_{e 1} \ p_{\nu 2}.p_{e 2} + p_{\nu 1}.p_{e 2} \ p_{\nu 2}.p_{e 1} \right\}  \nonumber \\
&+& 2  \left(\frac{8 G_F }{\sqrt{2} \ M_{\zp}^2}\right) \ u_l^{\nu} \  u_l^{e } \nonumber\\
&&\left\{ (c_v + c_a) \ p_{\nu 1}.p_{e 1} \ p_{\nu 2}.p_{e 2} \ \right. \nonumber\\ 
&& \hspace{2cm} \left. + \ (c_v - c_a)  \ p_{\nu 1}.p_{e 2} \ p_{\nu 2}.p_{e 1}\right\} \nonumber\\ 
&+& 2 \left(\frac{16 G_F }{\sqrt{2} \ M_{\zp}^2}\right) \ u_l^{\nu} \  u_l^{e }  \ 
p_{\nu 1}.p_{e 1} \ p_{\nu 2}.p_{e 2}  .\nonumber
\end{eqnarray}
where we neglected the electron mass and took $c_v' = c_v + 1$, $c_a' = c_a + 1$.

\vspace{0.3cm}
According to our previous analysis, one can fit CHARM II results by taking   
$u_l^{\nu} = u_l^e \sim [0.3,0.6] \ 10^{-6} (m_{\zp}/\mbox{MeV})$. With these couplings, 
the ``new'' $\nu_e e$ elastic scattering cross section then appears to be slightly larger than the Standard Model prediction 
(it would be slightly lower if $u_l^e < 0$). We then obtain
$$ \sigma_{\nu-e}^{sm} \sim [9.4, 9.8] \, 10^{-45} \left(\frac{E}{\rm{MeV}}\right) \rm{cm}^2.$$
The difference between our prediction and the Standard Model expectation is too small to be detected yet.

\vspace{0.3cm}
Finally, one could investigate the effect of a gauge boson lighter than $m_{\mu}$. 
The situation is quite identical to what is described above, except that it seems difficult 
to satisfy the neutrino experiment constraints and explain the NuTeV anomaly simultaneously.

\subsection{MUNU experiment}

Let us  now study the modification of the $\bar{\nu}_e e$ elastic scattering cross section. 
The latter is similar to $\nu_e e$, except that i) the $W$ exchange now 
proceeds through a $s$-channel (instead of a $t$-channel) and ii) the scalar products 
$p_{\nu 1}.p_{e 2} p_{\nu 2}.p_{e 1}$ and $p_{\nu 1}.p_{e 1} \ p_{\nu 2}.p_{e 2}$ are exchanged.

\vspace{0.3cm}
With the very small couplings mentioned above, we obtain that the new (extended) $\bar{\nu}_e e$ cross section is 
increased if $u_l^e >0$. It is decreased if $u_l^e < 0$. To our knowledge, no experiment measured 
$\sigma_{\bar{\nu}_e e}$ with great accuracy, apart from the MUNU experiment \cite{munu}. However, according to 
our estimates, this measurement does not allow one to set better constraints on the couplings of a light gauge boson. 
A proper experimental analysis is definitely required nevertheless.

\vspace{0.3cm}
An other experiment \cite{reines} also measured the $\bar{\nu}_e e$ cross section. 
They obtain a good fit to the 
Weinberg-Salam model by taking $\sw^2 = 0.29 \pm 0.05$. This value now turns out to be marginally compatible 
with the best fit of the Standard Model. Taken at face value, however, it tends to indicate that the $\bar{\nu}_e e$ 
cross section is higher than expected. This experimental result would favour the case $u_l^e >0$, and in fact 
would appear quite compatible with the presence of a light $\zp$.

\subsection{Back to the NuTeV anomaly}

Assuming $Q^2=20$ GeV$^2$ and $-u_l^{u}$ $= u_l^d$ $ =  u_l^{\nu} \sim [0.3, 0.6] \, \sqrt{1.5} \ 10^{-6} \ (m_{\zp}/\mbox{MeV})$,  we find that  
a gauge boson of $\sim$ 2-4 GeV could explain the anomaly very well (correcting $Q^2 \rightarrow Q^2+M_{\zp}^2$ in the formula of $R_{num}$ ). 
Relaxing universality and increasing the quark couplings by a factor 10, 
we obtain an excellent fit of the NuTeV anomaly by choosing $m_{\zp} \sim 546$ MeV (we then get $R_{num} = 3.1506$).

\vspace{0.3cm}
Taking $u_l^{u} <0$, $u_l^d > 0$, $u_l^e >0$ and $u_l^{\nu} >0$ could suggest that the 
$\zp$ couplings are proportional to $[T_3^f - 2 Q(f)]$ or $[T_3^f - 2 Q(f) \cos \theta]$ with $Q(f)$, $T_3^f$ the 
particle charge and isospin respectively and $\cos \theta > 0.75$. 
If $u_l^{u} = - u_l^d$ but $u_l^e = u_l^{\nu}$, then one could propose for example 
a different relationship between leptons and quarks, say $[T_3^f - 2 Q(f) \sin \theta]$ for leptons 
and $[T_3^f - 2 Q(f) \cos \theta]$ ($\cos \theta > 0.75$) although we doubt that this is realistic.
The previous remarks suggest that we should relax the universality assumption but we do not expect 
our conclusions to be changed drastically nevertheless.

\vspace{0.3cm}
The surprising agreement between our naive model and 
the NuTeV anomaly may just be a matter of coincidence. However, we think it may be worth taking a more careful look at  
the possibility of light gauge bosons, first because they seem to be still allowed by neutrino physics, 
secondly they may explain Standard Model deviations and finally 
they are expected to play a key role in the Light Dark Matter scenario\cite{bens,bf,raffelt,511,llbfs,fayet}.

 \vspace{0.3cm}
Whether such a gauge boson exists or not, we think that this result is of interest as it could perhaps 
motivate further analysis and experimental efforts to measure these cross sections (especially 
$\sigma_{\bar{\nu}_e e}$) more accurately.

\subsection{Neutrino oscillations}
Solar neutrino experiments measure the elastic scattering cross section $\nu_x + e^- \rightarrow \nu_x + e^-$, 
the neutral current cross section $\nu_x + d \rightarrow n + p + \nu_x$ and the charged current process 
$\nu_e + d \rightarrow p + p + e^-$. Adding a $\zp$ is likely to change both the elastic scattering and the 
neutral current cross section. However, given the couplings imposed by CHARM II, these modifications
are likely to be too small to be detected.  We estimate them to be less than a few percent, which seems under 
the sensitivity of solar and atmospheric neutrino experiments. 

\vspace{0.3cm}
A proper analysis is nevertheless required to constrain  the 
parameter space available for a new gauge boson from the precise measurement of 
neutrino oscillation parameters  \cite{oscillations}.

\subsection{Other deviations possibly related to a $\zp$}

Another type of experiment which indicates Standard Model deviations is related 
to atomic parity violation. The measured value of $Q_w$ for Cesium atoms slightly differs from theoretical 
predictions. A new gauge boson could perhaps do the job but this depends on the 
mixing angle between $\zp$ and $Z$ that is introduced. For our present analysis, we supposed 
no mixing. The mixing angle is also crucial for the determination of the modification of the 
$Z$ decay width and Left and Right asymmetries.

\section{Predictions}

A new experiment is proposed \cite{Imlay} to determine $\sw^2$ at 1 or 2 $\%$ accuracy at low energy ($\sim$ MeV). 
The idea is to measure   
$$ R= \frac{\sigma(\nu_{\mu} e)}{\sigma({\nu_e e}) + \sigma(\bar{\nu}_{\mu} e)} = 
\frac{0.75 - 3 \sw^2 + 4 \sw^4}{1+2 \sw^2 + 8\sw^4}.$$ 

\vspace{0.3cm}
For the Standard Model best fit at low energy, we find: $R=0.1438$. 
If their measurement of $\sw^2$ is $1 \%$ accurate, then this experiment should measure $R \in [0.1414, \, 0.1463]$.  With 
$2 \%$ accuracy, $R \in [0.1391, \, 0.1488]$.   

\vspace{0.3cm}
Our prediction for 
$u_l = 0.1 \, \sqrt{1.5} \, 10^{-6} \ (m_{\zp}/\mbox{MeV})$  is $R=0.1433$ (which is within $1 \%$ accuracy so this should not be detected). 
For $u_l = 0.35 \, \sqrt{1.5} \, 10^{-6} \ (m_{\zp}/\mbox{MeV})$,  $R=0.1382$ (slightly more than $2 \%$ which is 
within the sensitivity of the experiment). Finally, for $u_l = 0.6 \, \sqrt{1.5} \, 10^{-6} \ (m_{\zp}/\mbox{MeV})$, we obtain $R=0.1285$ 
(which is a $\sim 6.8 \%$ deviation of the 
expected value of $\sw$). 

\vspace{0.3cm}
This experiment should therefore be able to test the very small couplings we had to consider for fitting both CHARM II, LSND and 
the NuTeV anomaly, unless $u_l < 0.3 \sqrt{1.5} \, 10^{-6} \ (m_{\zp}/\mbox{MeV})$.

\vspace{0.3cm}
Note that we do not need to specify the mass of the gauge boson to make these predictions. 
This experiment being at low energy, the mass term $m_{\zp}$ that appears in the expression of the 
couplings cancel out with the propagators.

\section{Conclusion}

We studied the implications of a light gauge boson on neutrino physics in light of the NuTeV anomaly and 
the experimental value of the muon $g-2$. We found that a $\zp$mass of about $\sim$ 314 MeV could explain both of them. 
However, the couplings required to explain the anomalous value of the $g-2$ yields 
a too large contribution to the $\nu e$ elastic scattering cross section at low energy. 

\vspace{0.3cm}
A way out is to consider 
much lower couplings (say a factor 3-10 of difference). One is then able to fit the NuTeV anomaly and satisfy 
both LSND and CHARM II constraints. 

\vspace{0.3cm}
The mass of the gauge boson that would have such qualities is expected to 
be of a few GeV ($\sim 3.7$ GeV for $u_l \sim 0.3 \,  \sqrt{1.5} \, 10^{-6} \ (m_{\zp}/\mbox{MeV})$). 
It would be smaller if the discrepancy (about the value of $\sw$) between NuTeV and the Standard Model 
turns out to be reduced after determination of strange sea asymmetry or isospin violation effects. 

\vspace{0.3cm}
The contribution of such a gauge boson 
to the muon $g-2$ would be too small to explain the discrepancy between Standard Model expectations and 
the experimental result. However, this can be easily explained by the presence of other particles $F$.  
If the latter are heavy enough (and their couplings not too large), they could have escaped 
past neutrino and accelerator experiments but nevertheless explain the value of the muon $g-2$ determined by E821. 
Such $F$ particles should be found at LHC.

\vspace{0.3cm}
The scenario of a new light gauge boson supplemented by other particles  has already been invoked in 
the framework of Light Dark Matter \cite{bf}, as well as in supersymmetry \cite{pierreu1}.

\vspace{0.3cm}
We note however that a gauge boson heavier than a few GeV (and as weakly coupled as what we found to 
satisfy the neutrino experimental constraints) would be marginally compatible with 
the LDM scenario (unless one considers a Dark Matter mass of $\sim$ 100 MeV), as 
the gauge boson couplings to Dark Matter would be close to be in the non perturbative regime.

\vspace{0.3cm}
We obtain a good fit of the NuTeV anomaly by considering that the quark couplings obey the relationship 
$u_l^d= - u_l^u$ while in the lepton sector, fitting the 
results for the $\bar{\nu}_e e$ elastic scattering cross section would rather suggest 
$u_l^{\nu}= u_l^e$ (the relationship  $u_l^{\nu}= - u_l^e$ is not excluded though). 
This could perhaps indicated that the $\zp$ couplings are proportional to a combination of the 
particle's charge and isospin. 

\vspace{0.3cm}
Whether a light gauge boson exists or not, we find surprising that, at present, low energy experiments may still 
allow for such a possibility.  More constraints should be obtained nevertheless from the precise measurement of the 
neutrino magnetic moment and maybe the neutrino oscillation parameters.

\section{Ackowledgements}
The author would like to thank J. March-Russell and G. Myatt 
for illuminating discussions, A. Dedes and S. Davidson for 
very useful references as well as 
S. Biller,  C. Burgess, I. de la Calle, J.M. Frere,  T. Hambye, N. Jelley, W. Porod, J. Silk, K. Zuber for very helpful comments.  
My deeplest regards to C. Nicholls. C.B. is supported by an individual PPARC Fellowship.

\end{document}